\documentclass[%
 reprint,
amsmath,amssymb, aps,
]{revtex4-1}

\usepackage{graphicx}
\usepackage{dcolumn}
\usepackage{bm}
\usepackage{float}

\begin{document}
\preprint{APS/123-QED}

\title{Reconstructing nonlinear networks subject to fast-varying noises by using linearization with expanded variables}
\author{Rundong Shi$^{1}$, Gang Hu$^{2}$, Shihong Wang$^{1\ast}$}

\address{$^{1}$School of Sciences, Beijing University of Posts and Telecommunications,
Beijing 100876,China\\
$^{2}$Department of Physics, Beijing Normal University, Beijing
100875, China\\
$^{\ast}$Corresponding author shwang@bupt.edu.cn}
\date{\today}

\begin{abstract}
Reconstructing noisy nonlinear networks from time series of
output variables is a challenging problem, which turns to be very
difficult when nonlinearity of dynamics, strong noise impacts and
low measurement frequencies jointly affect. In this paper, we
propose a general method that introduces a number of nonlinear terms
of the measurable variables as artificial and new variables, and
uses the expanded variables to linearize nonlinear differential
equations. Moreover, we use two-time correlations to decompose
effects of system dynamics and noise impacts. With these
transformations, reconstructing nonlinear dynamics of original networks is approximately equivalent to
solving linear dynamics of the expanded system at the least squares approximations. We
can well reconstruct nonlinear networks, including all dynamic
nonlinearities, network links, and noise statistical
characteristics, as sampling frequency is rather low. Numerical
results fully justify the validity of theoretical derivations.
\end{abstract}

\maketitle
\section{Introduction}
Networks are investigated in many branches of science. During the last few decades, researchers
have shown quickly increasing interest in exploring network
structure from available output data, i.e., the so-called network reconstruction problem. They mainly focus on two
types of approaches: statistical methods and dynamical methods.
The statistical methods rely on simple linear correlations,
information entropy and statistical inferences, such as Pearson's
correlation coefficients \cite{1,2,3}, mutual information
\cite{4,5,6,7}, and Bayesian network network inferences \cite{8,9}.
Recently, various methods revealing network structures from
dynamic data have been also proposed, which are based on various
levels of pre-knowledge about systems. Yu et al proposed a
method of updating a network copy continuously until the
copy system exhibits a dynamics identical to the original system
\cite{10}. Timme proposed a driving-response approach to infer
network topology \cite{11}. Calculating derivatives of state
variables, Shandilya and Timme transformed differential equations
of systems into algebraic equations \cite{12}. They infered link
strengths by solving the over-determined algebraic equations
through minimal 2-norm. Wang et al depicted sparse network
structure by using compress sensors, which needed only small
amount of data for the network construction \cite{13}. Levnajic
and Pikovsky untangled links via derivative-variable correlations
\cite{14}, and so on.

In many cases, there exist noises in systems. Bayesian
inference has first opened the door to the analysis of
noisy systems \cite{15,16,17}. Correlation and
high-order correlation are used to treat
noisy systems \cite{18,19,20,21,22,23,24}. Considering
derivative-variable correlation, Zhang et al proposed an approach
inferring both network links and strengths of noises \cite{20}, and Chen et al developed
this method by using suitable bases to reconstruct all
dynamic nonlinearities, topological interaction links and noise
statistical structure \cite{22}, but this method very much requires high sampling frequency for computing derivatives of variables. Emily and Tam presented a method to reconstruct network links with a low measurement frequency by using
variable-variable correlation and variable-time-lagged-variable
correlation. This method is fairly accurate when the dynamics of each
node are around fixed points \cite{23}. Lai extended this
method to discrete-time dynamics \cite{24}. Recently, Stankovski et al reviewed five theoretical methods for the reconstruction of coupling functions and their applications in chemistry, biology, physiology, neuroscience, social sciences,
mechanics and secure communications \cite{25}.

There are various difficulties encountered in network
reconstruction: complexity of network structures; nonlinearity of
network dynamics; disturbances of noises; low data quality such as
low sampling frequency, and so on. Here, we present an approach to reconstruct
nonlinear networks subject to fast-varying noises from dynamic data only, including inferring
all nonlinearities and statistical noise structure. This method is
based on expanded variables and the least squares approximation.
Reconstructing nonlinear networks by computing expanded linear
networks is the novel feature of our approach, and its
good accuracy of inferring noisy nonlinear networks by
using low sampling frequencies is remarkable.
\section{Theroy}
Let us consider a general nonlinear dynamics subject to fast-varying noises
\vspace{-10pt}
\begin{eqnarray}
\dot{\textbf{x}}(t) &=&\textbf{f}(\textbf{x}(t))+\bm \eta(t),
\end{eqnarray}
where $\textbf{x}$ and $\bm \eta$ are the state vector
$\textbf{x}(t)=(x_{1}(t),x_{2}(t),...,x_{m}(t))^{T}$ and the noise
$\bm \eta(t)=(\eta_{1}(t),\eta_{2}(t),...,\eta_{m}(t))^{T}$,
where superscript $T$ denotes a transpose. Dynamic field reads
$\textbf{f} =(f_{1},f_{2},...,f_{m})^{T}$. Here we assume white
noise $\eta_{i}$ with zero mean and the following statistics
\vspace{-5pt}
\begin{equation}
\langle\eta_{i}(t)\rangle = 0,\langle\eta_{i}(t)\eta_{j}(t')
\rangle = D_{ij}\delta(t-t')
\end{equation}
with $i,j=1,2,...,m$. $D_{ij}=\sigma_{i}^{2}\delta_{ij}$. Our task is to depict nonlinearities $\textbf{f}$ and noise statistics $\textbf{D}$ from measurable data set $\textbf{x}(t_{1}),\textbf{x}(t_{2}),...,\textbf{x}(t_{N})$, $t_{i+1}-t_{i}=\tau$.

First we assume $f_{i}$s can be generally expanded by a basis set
as
\vspace{-10pt}
\begin{equation}
f_{i}(x) \approx \sum_{j=1}^{n}A_{ij}L_{j}(x)
\end{equation}
The basis set includes linear bases $L_{j}=x_{j}$ for $j=1,\
2,...,m$, nonlinear bases $L_{j}=h_{j}(\textbf{x}), m<j<n$, and
constant basis $1$. If measurement frequency is very high such that $\dot{\textbf{x}}(t)$ can be computed,
$A_{ij}$ can be inferred \cite{22}. Or, if Eq. (1) is approximately linear, $A_{ij}$ can be also inferred even when measurement frequency is rather low \cite{23}. In the following we will
show how to infer $A_{ij}$ when both
difficulties of nonlinear dynamics and low sampling frequency are
encountered.

The first key ingredient of our approach is that we transform nonlinear Eq. (1) to expanded linear differential equations. By taking nonlinear bases as new state variables and basing on Eqs. (1) and (3), we arrive at
\begin{equation}
\dot{\textbf{L}}(t)  = \textbf{A}\textbf{L}(t)+\textbf{R}(t)+\bm
\eta'(t)
\end{equation}
where $\textbf{A}\in \mathbb{R}^{n \times n}$ and
$\textbf{R}\in \mathbb{R}^{n \times 1}$ being the residual
vector due to limited bases. Obviously, the first $m$ rows of
$\textbf{A}$ is equal to the coefficient matrix of Eqs. (1) and (3).

Another key ingredient of our approach is to approximate $\textbf{R}(t)$ with the given bases $\textbf{L}(t)$, by using the least squares approximations, thus Eq. (4) is modified to
\begin{equation}
\dot{\textbf{L}}(t) = \textbf{B}\textbf{L}(t)+ \textbf{e}(t)+\bm
\eta'(t)
\end{equation}
where $ \textbf{e}(t)$ are the errors of approximation on an
interval $T_{1} \leq t\leq T_{2}$ and $e_{i}=0$ for $i=1,2,...,m$.
$ \textbf{e}(t)$ should satisfy the following formula
\begin{equation}
\frac{1}{T_{2}-T_{1}}\int_{T_{1}}^{T_{2}}\textbf{e}(s)\textbf{L}(s)^{T}ds =\langle \frac{\int_{t}^{t+\tau}\textbf{e}(s)\textbf{L}(s)^{T}ds}{\tau}\rangle = 0
\end{equation}
Due to $e_{i}=0$ for $i=1,2,...,m$, the first $m$ rows of
$\textbf{B}$ is equal to the first $m$ rows of $\textbf{A}$.

Now Eq. (5) becomes linear and its analytic solution can be given explicitly
\begin{equation}
\textbf{L}(t+\tau) = e^{\textbf{B}
\tau}\textbf{L}(t)+\int_{t}^{t+\tau}e^{\textbf{B}
(t+\tau-s)}(\textbf{e}(s)+\bm \eta'(s))ds
\end{equation}
Multiplying both sides of Eq. (7) by $\textbf{L}(t)^{T}$ and
averaging all the terms in the equation, we can obtain
\begin{eqnarray}
\langle\textbf{L}(t+\tau)\textbf{L}(t)^{T}\rangle &=& e^{\textbf{B}
\tau}\langle\textbf{L}(t)\textbf{L}(t)^{T}\rangle + \\
&& \langle\int_{t}^{t+\tau}e^{\textbf{B}
(t+\tau-s)}(\textbf{e}(s)+\bm \eta'(s)) \textbf{L}(t)^{T}ds\rangle \nonumber
\end{eqnarray}
where $\langle \bullet \rangle$ denotes averages of sampling data. Since
\begin{equation}
\langle\int_{t}^{t+\tau}e^{\textbf{B}
(t+\tau-s)}\textbf{e}(s) \textbf{L}(t)^{T}ds \rangle
\approx\langle \int_{t}^{t+\tau}\textbf{e}(s)\textbf{L}(s)^{T}ds\rangle=0
\end{equation}
and with time lag $\tau > 0$ noise-variable correlations
approximately vanish $
\langle\int_{t}^{t+\tau}e^{\textbf{B}(t+\tau-s)}\bm \eta'(s)
\textbf{L}(t)^{T}ds\rangle \approx 0 $, Eq. (8) can be reduced to
\begin{equation}
\langle\textbf{L}(t+\tau)\textbf{L}(t)^{T}\rangle =e^{\hat{\textbf{B}}\tau}\langle\textbf{L}(t)\textbf{L}(t)^{T}\rangle
\end{equation}
By defining $\textbf{S}_{\tau}=\langle\textbf{L}(t+\tau)\textbf{L}(t)^{T}\rangle$
and $\textbf{S}_{0}=\langle\textbf{L}(t)\textbf{L}(t)^{T}\rangle$, which are explicitly computable with the available data,
we rewrite Eq. (10) as
\vspace{-10pt}
\begin{equation}
\textbf{S}_{\tau}=e^{\hat{\textbf{B}} \tau}\textbf{S}_{0}
\end{equation}
Matrix $\hat{\textbf{B}}$ is thus solved as
\vspace{-5pt}
\begin{equation}
\hat{\textbf{B}}=\frac{\ln[\textbf{S}_{\tau}
\textbf{S}_{0}^{-1}]}{\tau}
\end{equation}
and the network reconstruction of the expanded linear differential equation (5)
is completed. With known $\hat{\textbf{B}}$, the original system
(1) is successfully inferred.

From the above analysis it is clear that our method typically
relies on variables expansion (Eq. (5)), then the task of inferring
nonlinear differential Eq. (1) is transformed into solving linear
differential equations. Noise effects are decorrelated by using
time-lagged correlation and the residuals of linearizing
$\textbf{R}$ can be projected on to the chosen bases by
using the least squares approximations (Eq. (5)). Basing on these transformations, we can
obtain the coefficient matrix by expanded variables and expanded
variable correlation matrices (Eq. (12)). Thus we name our method VELSA (variable expansion and least squares approximations)

We can also infer noise statistical matrix of Eq.
(2) from the available data. For $\textbf{L}(t+\tau) \approx
e^{\hat{\textbf{B}}
\tau}\textbf{L}(t)+\int_{t}^{t+\tau}e^{\hat{\textbf{B}}
(t+\tau-s)}\bm \eta'(s)ds$, multiplying its both sides by respective transposes and averaging all the terms, we obtain
\begin{widetext}
\begin{equation}
\langle\textbf{L}(t+\tau)\textbf{L}(t+\tau)^{T}\rangle = e^{\hat{\textbf{B}}\tau}\langle\textbf{L}(t)\textbf{L}(t)^{T}\rangle e^{\hat{\textbf{B}}^{T}\tau}
      +\langle \int_{t}^{t+\tau}e^{\hat{\textbf{B}}(t+\tau-s)}\bm\eta'(s)ds \int_{t}^{t+\tau}\bm\eta'(s')^{T}e^{\hat{\textbf{B}}^{T}(t+\tau-s')}ds'\rangle
\end{equation}
\end{widetext}
Based on Eqs. (2), (5) and (11), Eq. (13) can be reduced to
\begin{equation}
\textbf{S}_{0} -
e^{\hat{\textbf{B}}\tau}\textbf{S}_{0}e^{\hat{\textbf{B}}^{T}\tau}=
\int_{0}^{\tau}e^{\hat{\textbf{B}}(\tau-s)} \textbf{D}'
e^{\hat{\textbf{B}}^{T}(\tau-s)}ds
\end{equation}
where $D'_{ij}\delta(s-s')=\langle \eta'_{i}(s)\eta_{j}^{'}(s')^{T}\rangle$. We define the left hand side of Eq. (14) as
\begin{equation}
\textbf{F}(\tau)=\textbf{S}_{0} -
e^{{\hat{\textbf{B}}}\tau}\textbf{S}_{0}e^{{\hat{\textbf{B}}}^{T}\tau}
\end{equation}
Computing the first derivatives of $\textbf{F}(\tau)$, $\textbf{F}^{(1)}(\tau)=-e^{\hat{\textbf{B}}\tau}(\hat{\textbf{B}}\textbf{S}_{0}+\textbf{S}_{0}\hat{\textbf{B}}^{T})e^{\hat{\textbf{B}}\tau}$, we integrate the form above and have
\begin{equation}
\textbf{F}(\tau)-\textbf{F}(0)=-\int_{0}^{\tau}e^{\hat{\textbf{B}}\tau}(\hat{\textbf{B}}\textbf{S}_{0}+\textbf{S}_{0}\hat{\textbf{B}}^{T})e^{\hat{\textbf{B}}^{T}\tau}d\tau  \nonumber
\end{equation}
Then
\begin{equation}
\textbf{S}_{0} -
e^{{\hat{\textbf{B}}}\tau}\textbf{S}_{0}e^{{\hat{\textbf{B}}}^{T}\tau}=-\int_{0}^{\tau}e^{\hat{\textbf{B}}\tau}(\hat{\textbf{B}}\textbf{S}_{0}+\textbf{S}_{0}\hat{\textbf{B}}^{T})e^{\hat{\textbf{B}}^{T}\tau}d\tau
\end{equation}
Comparing Eqs. (14) and (16), we obtain an identical formula
\begin{equation}
\textbf{D}' =
-(\hat{\textbf{B}}\textbf{S}_{0}+\textbf{S}_{0}\hat{\textbf{B}}^{T})
\end{equation}
where the noise statistic matrix $\textbf{D}$ of Eq. (1) is a sub-matrix of
$\textbf{D}'$, $D'_{ij}=D_{ij}$ for $i,j=1,\ 2,...,m$.

Errors of the VELSA method can be well analyzed (Detailed analysis in APPENDIX). Considering the residual errors of expanded variables, i.e., $\textbf{e}(t)$ in Eq. (5), we rewrite Eq. (10) as
\begin{equation}
\textbf{S}_{\tau}=e^{\textbf{B}
\tau}\textbf{S}_{0}+\langle\int_{t}^{t+\tau}e^{\textbf{B}
(t+\tau-s)} \textbf{e}(s)\textbf{L}(t)^{T}ds\rangle
\end{equation}
Further Taylor expanding the integral term and using the least squares approximations, we finally obtain
\begin{equation}
\textbf{B}=\hat{\textbf{B}}+e^{\hat{\textbf{B}}\tau}\langle\textbf{e}(t)\textbf{L}^{(1)}(t)^{T}\rangle
\textbf{S}_{0}^{-1}\frac{\tau}{2} + O(\tau^{2})
\end{equation}
Considering specific $e_{i}(t)$ and $e^{\hat{\textbf{B}}\tau}=\textbf{I}+O(\hat{\textbf{B}}\tau)$, we further obtain the errors of reconstruction
\begin{equation}
B_{ij}-\hat{B}_{ij}=\left\{
\begin{aligned}
O(\tau^{2}),\ if \ e_{i}(t)=0 \\
O(\tau) ,\ otherwise
\end{aligned}
\right.
\end{equation}
On summary, due to $R_{i}= 0$ in Eq. (5), $i=1,2,...,m$, i.e., $e_{i}(t) = 0$, the errors of reconstructed coefficients of Eq. (12) are proportional to $\tau^{2}$ and can be quickly reduced by increasing measurement frequency.

\section{simulations and results}
\subsection{Lorenz system}
For justifying the VELSA method, we first consider the Lorenz system subject to fast-varying noises
\begin{eqnarray}
\dot{x} &=& \sigma(y-x)+\eta_{1}(t), \nonumber \\
\dot{y} &=& \rho x-xz-y+ \eta_{2}(t),   \\
\dot{z} &=& xy- \beta z +\eta_{3}(t), \nonumber
\end{eqnarray}
where $\sigma=10$, $\rho=28$, $\beta=2$, at which deterministic dynamics is chaotic. Moreover, all variables are affected by strong noises, simplified to
$\sigma_{i}^{2}=100,i=1,2,3$.

Figure 1(a) shows a chaotic and random trajectory of the noisy system.
We use Eqs. (12) and (17) to compute $\hat{\textbf{D}}'$ and
$\hat{\textbf{B}}$, and specify $\hat{\textbf{A}}$ and $\hat{\textbf{D}}$ from $\hat{\textbf{B}}$ and $\hat{\textbf{D}}'$, respectively, ($\hat{A}_{ij}=\hat{B}_{ij}$, $i=1,2,...,m$, $j=1,2,...,n$, $\hat{D}_{ij}=\hat{D}'_{ij}$, $i,j=1,2,...,m$). First, we should
select proper bases to expand field functions. Without knowing any
particular information about the field functions, we generally
choose power series as a basis set, by assuming the following bases $(L_{1},\ L_{2}, \ ... ,\ L_{n})^{T}$
with truncation $n$,
\begin{equation}
( x,\ y,\ z,\
x^{2},\ xy,\ xz,\ y^{2},\ yz,\ z^{2},\ x^{3},\
... ,\ 1)^{T}
\end{equation}
Calculating
$\textbf{S}_{\tau}$ and $\textbf{S}_{0}$ with available data, we obtain reconstruction results given in
Figs. 1(b-f). In Fig. 1(b), the reconstruction results
$\hat{A}_{ij}$ with truncations of the first order ($n=4$) and the three order ($n=20$) are plotted against those with the second
order ($n=10$). The results of $\hat{A}_{ij}(n=4)$
deviates considerably from ones of $\hat{A}_{ij}(n=10)$, but
satisfactory identity between $\hat{A}_{ij}(n=10)$ and
$\hat{A}_{ij}(n=20)$ is observed. According to the self-consistent
checking method \cite{22}, by increasing the number $n$ of tested unknown variables, the reconstruction parameters with small $n$ remain unchanged (saturated) and the small $n$ (here $n=10$) is concluded as a sufficient and satisfactory expansion. In Fig. 1(c), it is clearly shown that all plots of $\hat{A}_{ij}$ and $\hat{D}_{ij}$ computed at $n=10$ are around the diagonal line, justifying satisfactory reconstruction.

\begin{figure*}
\includegraphics[width=12cm]{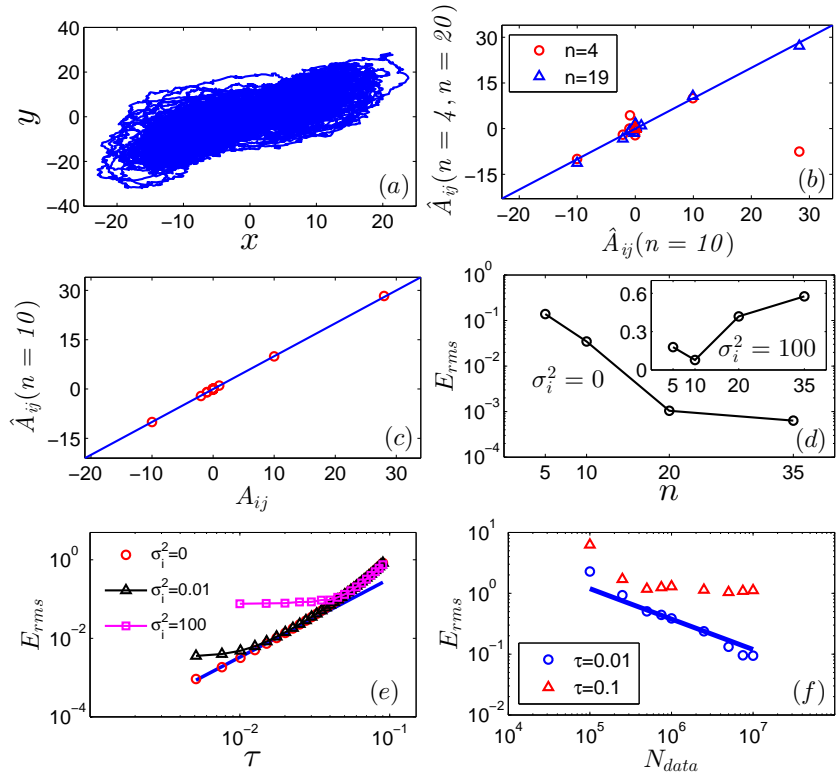}
\caption{\label{} Application of VELSA to Lorenz system. (a) Trajectories of noisy system with $\sigma_{i}^{2}=100$ and time step for simulation being
$10^{-4}$. (b-f) Reconstruction results with $n=10$, $N_{data}=10^{7}$ and $\tau=0.1$. (b) Reconstruction results $\hat{A}_{ij}$ with $n=4$
(the first order truncation) and $n=20$
(the third order truncation) plotted against those with $n=10$ (the second order truncation). (c) Reconstructed results
of $\hat{A}_{ij}$ and $\hat{D}_{ij}$ plotted against the actual coefficients. (d) Dependence of $E_{rms}$ on the basis number $n$ with noise $\sigma_{i}^{2}=0$ and $\sigma_{i}^{2}=100$ (inset).
(e) Dependence of $E_{rms}$ on $\tau$ with $\sigma_{i}^{2}=0,\ 0.01, \ 100$. (f) Dependence of $E_{rms}$ on the sampling number $N_{data}$ with $\tau=0.01$ and $\tau=0.1$. $\sigma_{i}^{2}=100$.}
\end{figure*}

To show the effects of bases, we calculate the root
mean square error
\begin{equation}
E_{rms}=\sqrt{\frac{\sum_{i=1}^{m}\sum_{j=1}^{n}(\hat{A}_{ij}-A_{ij})^{2}}{m\times n}},
\end{equation}
Figure 1(d)
presents $E_{rms}$  with noise $\sigma_{i}^{2}=0$ and $\sigma_{i}^{2}=100$ (inset). The bases of $n=5$
are the actual bases of Eq. (21), i.e., $x,\ y,\ z,\ xy,\ xz$. In Fig.
1(d) we observe that errors of the noise-free system
monotonically decrease with $n$, while due to
noise effects the results show an optimal and minimal error at about $n=10$ (small frame). Figure 1(e) shows the dependence of $E_{rms}$ on $\tau$ with $\sigma_{i}^{2}=0,\ 0.01,\
100$. We observe that results with
$\sigma_{i}^{2}=0$ (circles) well coincide with the line of $E_{rms}\propto
\tau^{2}$ for small $\tau$. For finite noises, errors decrease with the decrease of
$\tau$, however, the decreasing tendencies saturate at small $\tau$'s to finite errors depending on noise intensities. In Fig. 1(f) Error dependences of the
sampling number $N_{data}$ with $\tau=0.01$ (circles) and
$\tau=0.1$ (squares) are plotted. The results of $\tau=0.01$
monotonically decrease, approximately proportional to $\frac{1}{\sqrt{N_{data}}}$. However, errors for
$\tau=0.1$ tend to saturation, which is determined by low measurement frequency. The
$E_{rms}$ behaviors in both Figs. (e)(f) clearly show the different effects of
$\sigma_{i}$, $\tau$ and $N_{data}$, and the competitive rules of $N_{data}^{-\frac{1}{2}}$, $\sigma_{i}$ and $\tau^{2}$ for $E_{rms}$. For detailed software codes of the method, see Ref. \cite{26}.

\begin{figure*}
\includegraphics[width=14cm]{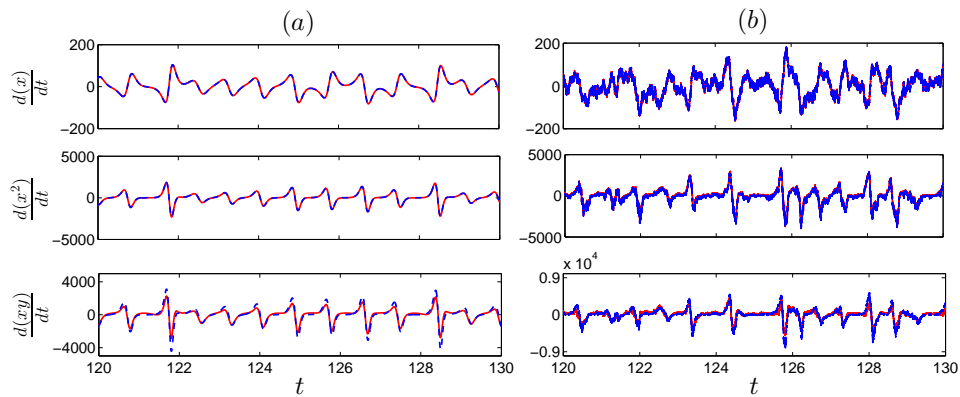}
\caption{\label{fig:epsart} Trajectories of Eq. (5) with no noise ((a)) and noise impacts $\sigma_{i}^{2}=100$ ((b)). $n=10$. Variables have no residual terms in their corresponding equations in top and middle panels while they do have in bottom panels. Solid-red and dash-blue curves are reconstructed and actual trajectories, respectively. Both types of curves coincide with each other perfectly when the corresponding equations have no residual terms, and small fluctuations are observed while nonzero residual terms exist.}
\end{figure*}
For examining the validity of Eq. (12) we compare the reconstruction trajectories of Eq. (5) by setting $e(t)=0$ and the actual ones in Fig. 2. It is shown that without noise (Fig. 2(a)) the two trajectories are almost identical when $\dot{L}_{i}(t)$ equations do not contain residual term while small deviations are observed when residual exists. With noises (Fig. 2(b)) the above conclusions are still valid, but reconstruction curves show some fluctuations caused by noises.

For demonstrating effectiveness of Eq. (12) we plot the trajectories of reconstruction system in Fig.3(a) and 3(b), corresponding to noisy and noise-free reconstruction systems, respectively. Figure 3(c) shows the trajectories of an original noise-free system. From comparison of Fig. 3(a) and Fig. 1(a), Fig. 3(b) and 3(c), we draw a conclusion that using our VELSA method not only reconstructs the noisy system, but also predicts the noise-free system.
\begin{figure}[htb]
\includegraphics[width=8cm]{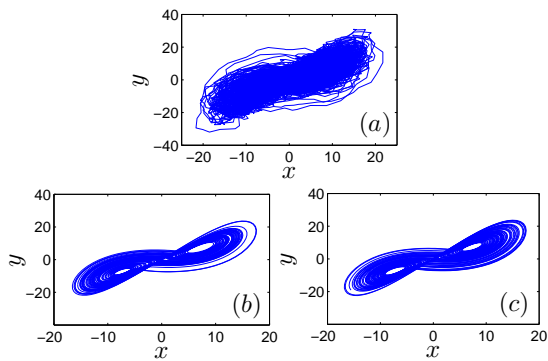}
\caption{\label{fig:epsart} Trajectories of Lorenz systems. Reconstruction system of an original noisy system with reconstructed noises (a) and without noises (b). Original noise impacts $\sigma_{i}^{2}=100$. (c) Actual noise-free system.}
\end{figure}

\subsection{A FHN neural network}
We now consider a more complicated nonlinear network, the
noisy FHN neural network,
\begin{eqnarray}
\dot{v_{i}} &=& \frac{1}{\epsilon}(v_{i}-\frac{1}{3}v_{i}^{3}-u_{i}+I)+\sum _{j=1}^{N}c_{ij}(v_{j}-v_{i})+\eta_{1}(t), \nonumber \\
\dot{u_{i}} &=& \gamma v_{i}-u_{i}+b+ \eta_{2}(t),
\end{eqnarray}
where we take $\epsilon=0.1$, $I=0$, $\gamma=b=1.5$ and $N=10$. Noise
$\sigma_{i}^{2}=0.1, i=1,\ 2$. $c_{ij}$ are coupling strengths generated from
the connection coefficients with connection probability $0.3$ and the
weighted coefficients uniformly distributed in $[0.4, \ 4.0]$. Here we separately define coupling coefficients as
$c_{ij}$, and local dynamic coefficients in Eq. (3) as $f_{ij}$,
$\dot{v}_{i}=f_{iv}(v,u)$, $\dot{u}_{i}=f_{iu}(v,u)$. We
can reconstruct $\hat{c}_{ij}$ and $\hat{f}_{ij}$ plotted in Figs. 4(a)(b), and
satisfactory identities are observed.
\begin{figure}[H]
\includegraphics[width=9cm]{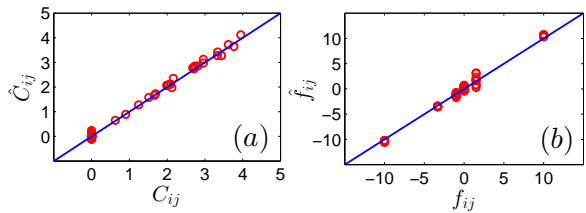}
\caption{\label{fig:epsart} Reconstruction of a 10-node FHN
network. Reconstructed $\hat{c}_{ij}$ and $\hat{f}_{ij}$ plotted
against actual $c_{ij}$ (a) and $f_{ij}$ (b). Variables are taken up to the third order of power expansion.
$\tau=0.1$. $N_{data}=4\times 10^{6}$.}
\end{figure}
\section{Comparison and Discussion}

After all the above demonstration of network reconstructions, a detailed comparisons between the VELSA method and the two previous methods are in order. Here the main problem in the present study is the joint difficulties of (i) Nonlinearity; (ii) Noise; (iii) Low measurement frequency. In [22], authors considered (i) and (ii) together with very fast data measurement such that velocities of $\textbf{x}(t)$ can be
accurately computed. In [23] authors considered (ii) and (iii) together by considering trajectories around a fixed point where network dynamics can be directly treated with linear approximations. Both methods fail if all difficulties of (i), (ii) and (iii) appear together.In Fig. 5(a), we compare the results of reconstruction of Eq. (21). While VELSA shows low $E_{rms}$ for rather wide $\tau$ range, the method HOCC in \cite{22} produces large errors. In Fig. 5(b) we compare the results of reconstruction of Eq. (24), and it is clearly shown that while VELSA satisfactorily infers all the network links, the method in \cite{23} fails to do so.
\begin{figure}[H]
\includegraphics[width=9.5cm]{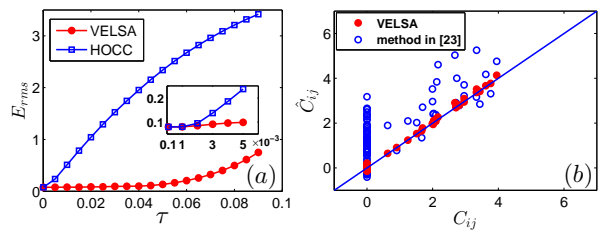}
\caption{\label{fig:epsart} Comparison of three methods. (a) Dependence of $E_{rms}$ on
 $\tau$ for HOCC \cite{22} and VELSA. (b) Calculated results $\hat{c}_{ij}$ plotted against the actual coefficients
$c_{ij}$ for the method in \cite{23} and VELSA.}
\end{figure}

For treating reconstruction problem of nonlinear and noisy network Eq. (1), the VELSA method (i) Transfers nonlinear terms of network Eq. (1) to expanded variables of equivalent linear network Eq. (4) while it is not closed; (ii) Uses the least squares approximations Eq. (6) to close the expanded linear network Eq. (4), i.e., to derive Eq.(5); (iii) Analytically solves Eq. (5) through two-time correlations by decomposing effects of system dynamics and noise impacts. Numerical computations fully justify the least squares approximations (Fig. 2), and the efficiency of reconstruction computation Eq. (12) (Figs. 1(b)(c), Fig.3 and Fig. 4).

With the above method we can achieve the following: (i) Though the reconstruction is applied for purely linear network Eq. (5), it fully includes strongly nonlinear effects (see Figs. 2, 3 and 4) due to that all nonlinear terms are taken into account in different expanded variables of Eq.(5) (in variables of $n\geq m$), and this is essentially different from linearization in conventional sense (e.g., see the comparison in Fig. 5(b)); (ii) Due to the utilization of analytical solution of linearized network Eq. (12), this method can obtain much better results when the measurement frequency is relatively low (see the analysis of Eq. (A10) and the comparison in Fig. 5(a)); (iii) The VELSA method can infer with measurable data not only the network structure in Eq. (12), but also noise statistical matrix $\textbf{D}$  in Eq. (17). Then with the reconstructed network we can predict the behaviors of the original network without subjecting noises (see Figs. 3(b)(c)), and reproduce the behaviors of the actual network under the impacts of realistic noises (compare Figs.1(a) and 3(a)).

The proposed method can be applied in real systems to infer network structures under certain conditions. At the present stage the VELSA method has its own limitations restricting the practical applications of the method. First, we consider white noise approximations for treating fast-varying noises. Extensions to slow-varying noises or even to noises with wide spectra should be further investigated. Second, this method usually includes large number of unknown parameters to be reconstructed, and thus need large data sets in computations, and small data sets can cause large errors (see circles in Fig. 1(f)). How to improve reconstruction precision when data sets are relatively small is still an important and unsolved problem. Third, the VELSA method can well treat data collected with much lower measurement frequency, in comparison with all the methods where time-derivatives from data are needed for reconstructions (Fig. 5(a)). However, this capability is limited either. By decreasing measurement frequency $\frac{1}{\tau}$, the reconstruction errors increase and the method completely fail at very large $\tau$ (see triangles in Fig. 1(f)). Finally, in this paper we consider available data of all nodes in the dynamical network under investigation. The extension of the VELSA method to the cases with some nodes hidden is another subject of practical importance.

\section{Conclusion}
In conclusion, we have proposed a method to reconstruct noisy nonlinear networks with fairly low measurement frequency, including all dynamic nonlinearities, network links, and noise statistical characteristics. Our method linearizes the original nonlinear equations by using expanded variables and solving nonlinear dynamics becomes equivalent to solving linear dynamics at the least squares approximations. Numerical results fully verify the validity of theoretical derivations and error analysis.

\appendix*
\section{ERROR DERIVATION}
\textbf{Preliminaries}. The exponential of a matrix $\textbf{A}$ is defined by
\begin{equation}
e^{\textbf{A}}=\sum _{k=0}^{\infty}\frac{1}{k!}\textbf{A}^{k}.   \tag{A.1}
\end{equation}
The logarithm of a matrix $\textbf{A}$ is defined by
\begin{equation}
\ln(\textbf{I}-\textbf{A})=\sum _{k=1}^{\infty}\frac{(-1)^{k}}{k}\textbf{A}^{k}.  \tag{A2}
\end{equation}
where $\textbf{I}$ is an identity matrix and the eigenvalues of $\textbf{A}$ satisy the form $|\lambda|< 1$.

Taylor expansion for an integration is defined by
\begin{equation}
\begin{aligned}
\int_{t}^{t+\tau} F(s)L(s)ds=F(t)L(t)\tau+F^{(1)}(t)L(t)\frac{\tau^{2}}{2} \\
+F(t)L^{(1)}\frac{\tau^{2}}{2}+O(\tau^{3})
\end{aligned}    \tag{A3}
\end{equation}
where $F^{(1)}(t)$ and $L^{(1)}(t)$ are the first derivatives of $F(t)$ and $L(t)$ against $t$, respectively.

\textbf{Error derivation}. Considering the residual errors of expanded variables, i.e., $\textbf{e}(t)$, we have
\begin{equation}
\textbf{S}_{\tau}=e^{\textbf{B}
\tau}\textbf{S}_{0}+\langle\int_{t}^{t+\tau}e^{\textbf{B}
(t+\tau-s)} \textbf{e}(s)\textbf{L}(t)^{T}ds\rangle   \nonumber
\end{equation}
With right multiplying $\textbf{S}_{0}^{-1}$, the above formulas yields
\begin{equation}
\textbf{S}_{\tau}\textbf{S}_{0}^{-1}=e^{\textbf{B}
\tau}+\langle\int_{t}^{t+\tau}e^{\textbf{B}
(t+\tau-s)} \textbf{e}(s)\textbf{L}(t)^{T}ds\rangle\textbf{S}_{0}^{-1}  \tag{A4}
\end{equation}
Define
\begin{equation}
\textbf{C}=\langle\int_{t}^{t+\tau}e^{\textbf{B}
(t+\tau-s)} \textbf{e}(s)\textbf{L}(t)^{T}ds\rangle\textbf{S}_{0}^{-1},  \tag{A5}
\end{equation}
then $\textbf{S}_{\tau}\textbf{S}_{0}^{-1}=e^{\textbf{B}\tau}+\textbf{C}$. We transform it into
\begin{equation}
\textbf{B}\tau=\ln(\textbf{S}_{\tau}\textbf{S}_{0}^{-1}- \textbf{C}). \tag{A6}
\end{equation}

First we compute the integration term in $\textbf{C}$, $\int_{t}^{t+\tau}e^{\textbf{B} (t+\tau-s)} \textbf{e}(s)ds$. By using Eq. (A3), Taylor expansion yields

\begin{eqnarray}
\int_{t}^{t+\tau}e^{\textbf{B} (t+\tau-s)} \textbf{e}(s)ds=e^{\textbf{B}\tau}\textbf{e}(t)\tau-\textbf{B}e^{\textbf{B}\tau}\textbf{e}(t)\frac{1}{2}\tau^{2} \nonumber \\
+e^{\textbf{B}\tau}\textbf{e}^{(1)}(t)\frac{1}{2}\tau^{2}+O(\tau^{3}) \nonumber
\end{eqnarray}

and Eq. (A5) becomes
\begin{widetext}
\begin{equation}
\textbf{C}=e^{\textbf{B}\tau}[\langle\textbf{e}(t)\textbf{L}(t)^{T}\rangle\tau-\textbf{B}\langle \textbf{e}(t)\textbf{L}(t)^{T}\rangle\frac{\tau^{2}}{2}+\langle \textbf{e}^{(1)}(t)\textbf{L}(t)^{T}\rangle\frac{\tau^{2}}{2}+ O(\tau^{2})]      \tag{A7}
\end{equation}
\end{widetext}
By using Eq. (A3), we further arrive at
\begin{widetext}
\begin{equation}
\langle\int_{t}^{t+\tau}\textbf{e}(s)\textbf{L}(s)^{T}ds\rangle=\langle\textbf{e}(t)\textbf{L}(t)^{T}\rangle\tau+\langle\textbf{e}^{(1)}(t)\textbf{L}(t)^{T}\rangle\frac{\tau^{2}}{2}+\langle\textbf{e}(t)\textbf{L}^{(1)}(t)^{T}\rangle\frac{\tau^{2}}{2}+O(\tau^{3}) \nonumber
\end{equation}
\end{widetext}
Due to the least squares approximations
\begin{equation}
\frac{1}{T_{2}-T_{1}}\int_{T_{1}}^{T_{2}}\textbf{e}(s)\textbf{L}(s)^{T}ds =\langle \frac{\int_{t}^{t+\tau}\textbf{e}(s)\textbf{L}(s)^{T}ds}{\tau}\rangle = 0,  \nonumber
\end{equation}
we have
\begin{eqnarray}
\langle\textbf{e}(t)\textbf{L}(t)^{T}\rangle\tau+\langle\textbf{e}^{(1)}(t)\textbf{L}(t)^{T}\rangle\frac{\tau^{2}}{2} \nonumber\\
=-\langle\textbf{e}(t)\textbf{L}^{(1)}(t)^{T}\rangle\frac{\tau^{2}}{2}+O(\tau^{3}) \nonumber
\end{eqnarray}

and
\begin{equation}
\langle\textbf{e}(t)\textbf{L}(t)^{T}\rangle = O(\tau) \nonumber
\end{equation}
Basing on the above two transformations, we rewrite $\textbf{C}$ as
\begin{equation}
\textbf{C}=-e^{\textbf{B}\tau}\langle\textbf{e}(t)\textbf{L}^{(1)}(t)^{T}\rangle\frac{\tau^{2}}{2}\textbf{S}_{0}^{-1}+ O(\tau^{3}) \tag{A8}
\end{equation}
Further transforming Eq. (A6) into $\textbf{B}\tau=\ln(\textbf{I}-(\textbf{I}-\textbf{S}_{\tau}\textbf{S}_{0}^{-1}+\textbf{C}))$, considering small $\tau$, we have $\textbf{S}_{\tau}\textbf{S}_{0}^{-1}\approx \textbf{I}$. Together with Eq. (A2), we can derive
\begin{widetext}
\begin{eqnarray}
\ln(\textbf{I}-(\textbf{I}-\textbf{S}_{\tau}\textbf{S}_{0}^{-1}+\textbf{C}))&=&\sum _{n=1}^{\infty}\frac{(-1)^{n}}{n}(\textbf{I}-\textbf{S}_{\tau}\textbf{S}_{0}^{-1}+\textbf{C})^{n} \nonumber \\
&=&\sum _{n=1}^{\infty}\frac{(-1)^{n}}{n}(\textbf{I}-\textbf{S}_{\tau}\textbf{S}_{0}^{-1})^{n}-\textbf{C}+\frac{1}{2}(\textbf{I}-\textbf{S}_{\tau}\textbf{S}_{0}^{-1})\textbf{C}+\frac{1}{2}\textbf{C}(\textbf{I}-\textbf{S}_{\tau}\textbf{S}_{0}^{-1})+O(\textbf{C}^{2}) \nonumber \\
&=& \ln\textbf{S}_{\tau}\textbf{S}_{0}^{-1}-\textbf{C}+\frac{1}{2}(\textbf{I}-\textbf{S}_{\tau}\textbf{S}_{0}^{-1})\textbf{C}+\frac{1}{2}\textbf{C}(\textbf{I}-\textbf{S}_{\tau}\textbf{S}_{0}^{-1})+O(\textbf{C}^{2}) \nonumber
\end{eqnarray}
\end{widetext}
Since $\textbf{C}= O(\tau^{2})$ (Eq. (A8)) and $(\textbf{I}-\textbf{S}_{\tau}\textbf{S}_{0}^{-1})= O(\tau)$ (Eq. (A4)), we have
\begin{equation}
\textbf{B}\tau=\ln\textbf{S}_{\tau}\textbf{S}_{0}^{-1}-\textbf{C}+O(\tau^{3}) \tag{A9}
\end{equation}
Substituting Eq. (A8) into Eq. (A9), we finally obtain the error of reconstruction
\begin{equation}
\textbf{B}-\hat{\textbf{B}}= e^{\textbf{B}\tau} \langle\textbf{e}(t)\textbf{L}^{(1)}(t)^{T}\rangle \textbf{S}_{0}^{-1} \frac{\tau}{2}+ O(\tau^{2}). \tag{A10}
\end{equation}
where reconstructed matrix $\hat{\textbf{B}}=\frac{\ln(\textbf{S}_{\tau}\textbf{S}_{0}^{-1})}{\tau}$.

\end{document}